**Perturbations to aquatic photosynthesis due to high-energy cosmic ray induced muon flux in the extragalactic shock model.**

[1]Lien Rodriguez, [1]Rolando Cardenas and [2]Oscar Rodriguez

[1]Planetary Science Lab. Department of Physics. Universidad Central "Marta Abreu" de Las Villas. Santa Clara. Cuba

[2] Instituto Superior de Tecnologías y Ciencias Aplicadas. Havana. Cuba

**Abstract** We modify a mathematical model of photosynthesis to quantify the perturbations that high energy muons could make on aquatic primary productivity. Then we apply this in the context of the extragalactic shock model, according to which Earth receives an enhanced dose of high-energy cosmic rays when it is at the galactic north. We obtain considerable reduction in the photosynthesis rates, consistent with potential drops in biodiversity.

**I Introduction**

In the long course on Earth's biological evolution, several astrophysical phenomena might have delivered important doses of high energy muons on the planet's surface [Dar, Laor and Shaviv 1998; Atri and Melott 2011]. Most studies acknowledge the high penetration power of these muons, quoting that they can travel hundreds of meters through the ocean water column. However, the investigation of biological damage of muons on ocean phytoplankton is to be done. Therefore, in this paper we present the diminution on phytoplankton photosynthesis that a flux of high energy muons would do. We examine the scenario of the extragalactic shock model, according to which Earth receives an enhanced dose of high-energy cosmic rays when it is at the galactic north [Atri and Melott 2011].

**II Materials and methods**

The so called *E* model for photosynthesis [Fritz et al 2008], which uses irradiances *E* instead of fluences *H*, allows calculating the photosynthesis rate *P* and depth *z* in the ocean, normalized to the maximum possible photosynthesis rate $P_S$:

$$\frac{P}{P_S}(z) = \frac{1 - e^{-E_{PAR}(z)/E_S}}{1 + E_{UV}^*(z)} \qquad (1)$$

Where $E_{PAR}(z)$ is the irradiance of photosynthetically active radiation (PAR) at depth *z*, $E_{UV}^*(z)$ is the irradiance of (inhibitory) ultraviolet radiation at depth *z*, convolved with a biological spectrum (the reason for the asterisk), and $E_S$ is a parameter accounting for the efficiency of the species in the use of PAR. In this work, we do not consider the effect of the enhanced solar ultraviolet irradiation due to the potential depletion of the ozone layer when Earth is at the galactic north. Concerning radiation damage, this would be a minor effect compared to the influence of muons [Melott et al 2010]. Thus, we assume the current annual ground level average of solar irradiation at three different latitudes (0, 30 and 60 degrees) and then propagate this spectrum down the water column, using Lambert – Beer's law of Optics:

$$E(\lambda, z) = E(\lambda, 0^-) e^{-K(\lambda) z} \qquad (2)$$

In the above expression, the attenuation coefficients $K(\lambda)$ define the optical ocean water type [Peñate et al 2010]. Irradiances $E(\lambda, 0^-)$ just below the water surface are obtained after subtracting the reflected light:

$$E(\lambda, 0^-) = [1-R] E(\lambda, 0^+), \qquad (3)$$

where $R$ is the reflection coefficient (calculated with the Fresnel formulae), and $E(\lambda, 0^+)$ are the spectral irradiances just above the water surface.

Total irradiances are obtained through:

$$E_{PAR}(z) = \sum_{400nm}^{700nm} E(\lambda, z) \Delta \lambda \qquad (4)$$

$$E_{UV}^*(z) = \sum_{280nm}^{400nm} \varepsilon(\lambda) E(\lambda, z) \Delta \lambda, \qquad (5)$$

with $\varepsilon(\lambda)$ being the biological action spectrum for photosynthesis inhibition under the action of the ultraviolet radiation. For the latitude 60 degrees, we use the same action spectrum as in [Cockell 2000]. For the latitudes 0 and 30 degrees, we use a biological action spectrum more adequate for temperate phytoplankton [Avila, Cardenas and Martin 2012]. In general, biological action spectra quantify the biological effects of electromagnetic radiation, giving more weight to those wavelengths more harmful. In the case of the ultraviolet bands considered in this work (UV-B: from 280 to 320nm, and UV-A: from 320 to 400nm), the values of the spectrum are higher for the former, not only because of more energetic photons, but also due to increased quantum absorption probabilities.

The $E$ model for photosynthesis (eq. (1)) was developed and tested under the ordinary background of ionizing radiation on current Earth. To account for important fluctuations of ionizing radiations we modify it to:

$$\frac{P}{P_S}(z) = \frac{1 - e^{-E_{PAR}(z)/E_S}}{f_{ir}(z) + E_{UV}^*(z)}, \qquad (6)$$

where $f_{ir}(z)$ is some sort of normalized dose of absorbed ionizing radiation at ocean depth $z$. In this work we focus on an scenario in which muons are the dominant contribution to biological damage, due to their high penetration power on ocean water. Studies on biological damage of muons on non-human samples are scarce [Atri and Melott 2011]. However, some studies suggest that doses are proportional to the overall muon flux, and that the the fluence-to-dose factor has little variation with energy [Chen 2006; Ferrari, Pelliccioni & Pillon 1997; Pelliccioni 2000; Sato, Endo & Niita 2011]. Particularly, Fig. 2 of [Sato, Endo & Niita 2011] shows that effective dose conversion coefficients for high energy muons (energy range $10^2$ - $10^5$ MeV) have a very soft dependence with muon energy. References mentioned in this paragraph led us to accept, as in [Atri and Melott 2011], the *ansatz* that for the case or irradiation with high energy muons, the enhanced dose $D_{enh}$ at Earth's ground would be proportional to the enhanced muon flux $F_{enh}$:

$$\frac{F_{enh}}{F_n} = \frac{D_{enh}}{D_n} \qquad (7)$$

The subscript $n$ refers to the respective magnitudes during the ordinary radiation regime. We then propose as the normalized dose $f_{ir}(0)$ of ionizing radiation at ground level:

$$f_{ir}(0) \equiv \frac{D_{enh}}{D_n} = \frac{F_{enh}}{F_n} \qquad (8)$$

In this work we consider $f_{ir}(z)$ constant down the water column. The reasons for this are that this function is a ratio rather than an absolute magnitude and that, due to light availability, photosynthesis is basically performed only in the first 200 meters of the water column.

With the considerations above, the modified model for photosynthesis stands:

$$\frac{P}{P_S}(z) = \frac{1 - e^{-E_{PAR}(z)/E_S}}{f_{ir}(0) + E_{UV}^*(z)} \qquad (9)$$

This is a first attempt to quantify the effects of muons in photosynthesis, which could be refined in future studies. On another hand, notice that when there is not deviation of the ionizing radiation flux from the average radiation background, $f_{ir}(0) = 1$, and we get the original $E$ model (eq. (1)).

In [Atri and Melott 2011], two extreme cases are considered, when Earth is at galactic north:

1) Minimum enhancement of ionizing radiation: $f_{ir}(0) = 1,26$

2) Maximum enhancement of ionizing radiation: $f_{ir}(0) = 4,36$

# III Results and discussion

For the sake of brevity, in Figures 1-6 we only show the photosynthesis rates for the case of the maximum enhancement of radiation. Information on the scenario with minimum enhancement of radiation is compacted in Tables 1 and 2.

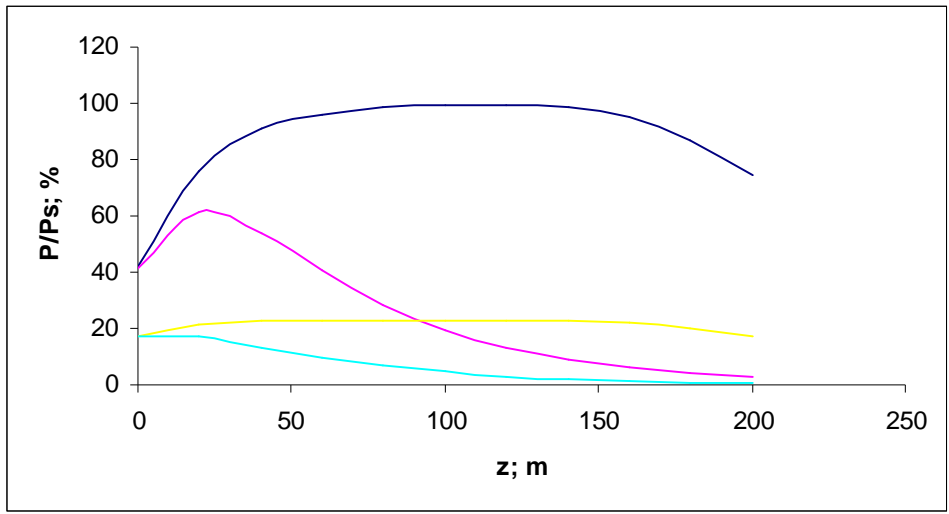

**Fig. 1** The two lower (at the start) curves show the photosynthesis rates for the maximum enhancement of ionizing radiation, for highly ($E_S = 2$ W/m$^2$) and poorly ($E_S = 100$ W/m$^2$) efficient organisms in the use of light. Latitude 0 degrees and optical ocean water type I. Upper curves represent the rates for ordinary conditions, for the sake of comparison.

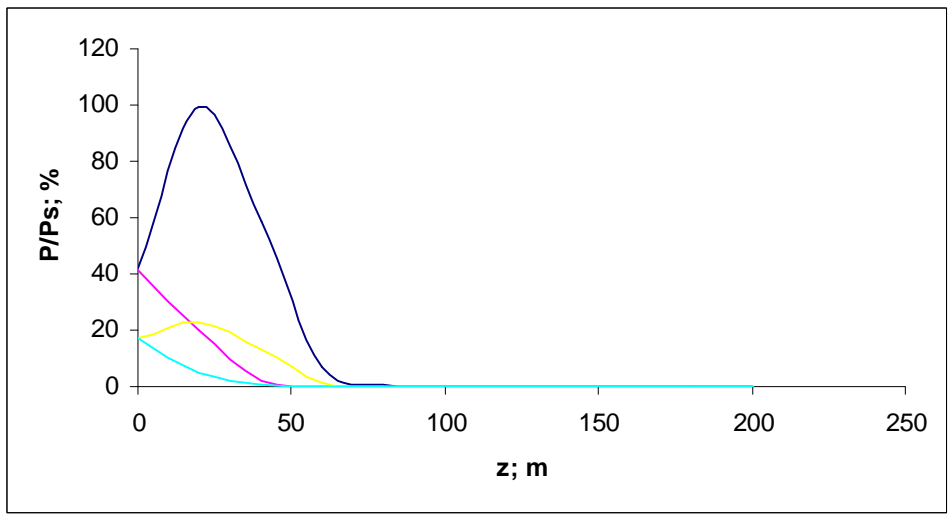

**Fig. 2** The two lower (at the start) curves show the photosynthesis rates for the maximum enhancement of ionizing radiation, for highly ($E_S = 2$ W/m$^2$) and poorly ($E_S = 100$ W/m$^2$) efficient organisms in the use of light. Latitude 0 degrees and optical ocean water type III. Upper curves represent the rates for ordinary conditions, for the sake of comparison.

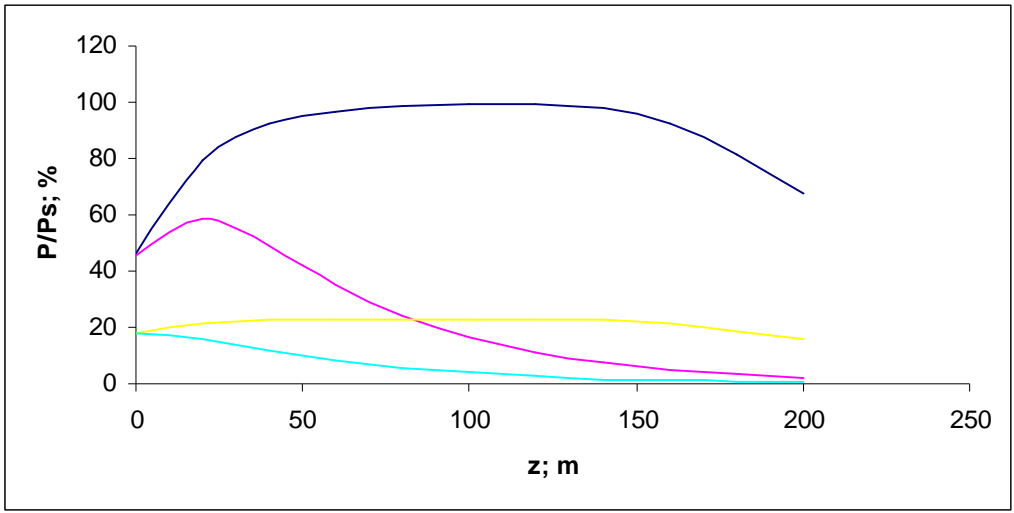

**Fig. 3** The two lower (at the start) curves show the photosynthesis rates for the maximum enhancement of ionizing radiation, for highly ($E_S = 2$ W/m$^2$) and poorly ($E_S = 100$ W/m$^2$) efficient organisms in the use of light. Latitude 30 degrees and optical ocean water type I. Upper curves represent the rates for ordinary conditions, for the sake of comparison.

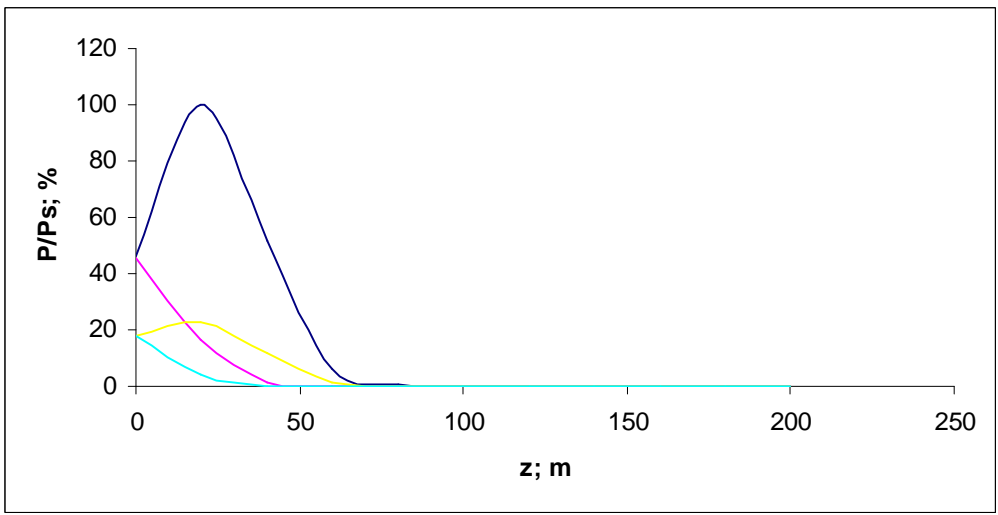

**Fig. 4** The two lower (at the start) curves show the photosynthesis rates for the maximum enhancement of ionizing radiation, for highly ($E_S = 2$ W/m$^2$) and poorly ($E_S = 100$ W/m$^2$) efficient organisms in the use of light. Latitude 30 degrees and optical ocean water type III. Upper curves represent the rates for ordinary conditions, for the sake of comparison.

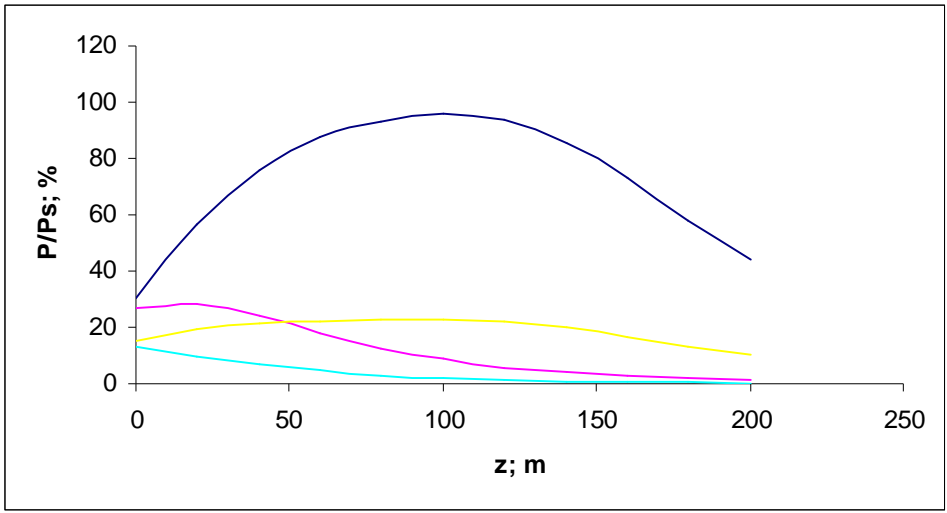

**Fig. 5** The two lower (at the start) curves show the photosynthesis rates for the maximum enhancement of ionizing radiation, for highly ($E_S$ = 2 W/m$^2$) and poorly ($E_S$ = 100 W/m$^2$) efficient organisms in the use of light. Latitude 60 degrees and optical ocean water type I. Upper curves represent the rates for ordinary conditions, for the sake of comparison.

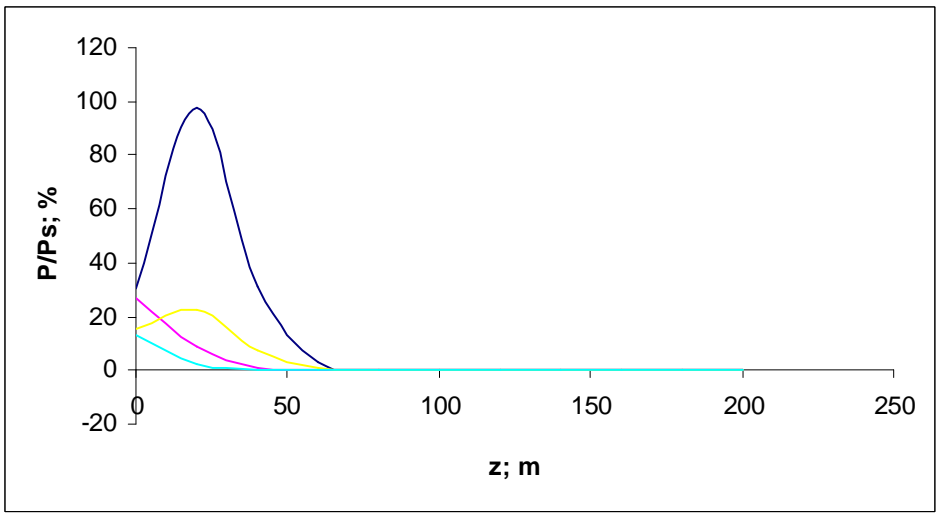

**Fig. 6** The two lower (at the start) curves show the photosynthesis rates for the maximum enhancement of ionizing radiation, for highly ($E_S$ = 2 W/m$^2$) and poorly ($E_S$ = 100 W/m$^2$) efficient organisms in the use of light. Latitude 60 degrees and optical ocean water type III. Upper curves represent the rates for ordinary conditions, for the sake of comparison.

Above plots show great diminution of photosynthesis rates in all cases. However, to get a more quantitative idea of this, we show in Table 1 the average photosynthesis rates in the photic zone (where light availability still allows for significant photosynthesis rates, here taken to be from the surface down to 200 meters). Then, in Table 2 it is shown the relative reduction of the average photosynthesis rates in the photic zone, for both the minimum and maximum enhancement of ionizing radiation.

| Ionizing Radiation Regime | Latitude (degrees) | Ocean Optical Water Type | Average Photosynthesis Rates <P/Ps>; % | |
|---|---|---|---|---|
| | | | Es=2W/m² | Es=100W/m² |
| Normal | 0 | I | 90 | 25,8 |
| | | III | 18,7 | 4,26 |
| | 30 | I | 89,4 | 23,4 |
| | | III | 18,1 | 4,13 |
| | 60 | I | 75,6 | 12,0 |
| | | III | 14,7 | 2,3 |
| Minimum enhancement | 0 | I | 72,2 | 21,1 |
| | | III | 15,1 | 3,6 |
| | 30 | I | 71,7 | 19,1 |
| | | III | 14,6 | 3,5 |
| | 60 | I | 61,3 | 10,1 |
| | | III | 11,9 | 2,0 |
| Maximum enhancement | 0 | I | 21,7 | 6,75 |
| | | III | 4,68 | 1,37 |
| | 30 | I | 21,5 | 6,1 |
| | | III | 4,5 | 1,3 |
| | 60 | I | 10,1 | 0,3 |
| | | III | 3,8 | 0,9 |

**Table 1** Average photosynthesis rates for the normal and radiation-enhanced scenario

| Ionizing Radiation Regime | Latitude (degrees) | Ocean Optical Water Type | Relative Variation of Average Photosynthesis Rates <P/Ps>; % | |
|---|---|---|---|---|
| | | | Es=2W/m$^2$ | Es=100W/m$^2$ |
| Minimum enhancement | 0 | I | 80,2 | 81,8 |
| | | III | 80,7 | 84,5 |
| | 30 | I | 80,2 | 81,6 |
| | | III | 80,7 | 84,7 |
| | 60 | I | 81,1 | 84,2 |
| | | III | 81,0 | 87,0 |
| Maximum enhancement | 0 | I | 24,1 | 26,2 |
| | | III | 25,0 | 32,2 |
| | 30 | I | 24,0 | 26,1 |
| | | III | 22,7 | 31,5 |
| | 60 | I | 13,4 | 2,5 |
| | | III | 25,9 | 39,1 |

**Table 2** Relative reduction of average photosynthesis rates (as compared to the normal radiation scenario)

In the case of minimum enhancement of radiation, for all latitudes the reduction in photosynthesis rates is around 20%. For the case of the maximum enhancement, the reduction of photosynthesis rates is drastic: they drop from 4 to 5 times and in some cases even more. This could be a factor originating an important descent in biodiversity,

**IV Conclusions**

If the periodic position of Earth at the north of the galaxy implies enhancements of ionizing radiation as those presented in [Atri and Melott 2011], then a considerable drop of phytoplankton photosynthesis is to be expected, especially when enhancement is close to the maximum values shown in the above reference. Being phytoplankton the starting point of the food assemblage, such a perturbation in its photosynthesis could cause a considerable drop in biodiversity, reinforcing the hypothesis in [Atri and Melott 2011] on a periodicity of around 62 My in fossil biodiversity. However, we point out that our conclusions are based only on the influence of radiations on photosynthesis. Other environmental variables (especially temperature) could change with an increase of the flux of high energy ionising radiations on the top of the atmosphere. Therefore, a more complete model of photosynthesis (in general, of habitability) is needed to improve the theoretical assessment for potential drops in biodiversity when Earth is near the north of our galaxy. This is ongoing work in our group.